

%
%
%
\def\unredoffs{} \def\redoffs{\voffset=-.31truein\hoffset=-.59truein}
\def\speclscape{\special{ps: landscape}}
%
%
%
%
\newbox\leftpage \newdimen\fullhsize \newdimen\hstitle \newdimen\hsbody
\tolerance=1000\hfuzz=2pt
\catcode`\@=11 
\def\bigans{b }
\message{ big or little (b/l)? }\read-1 to\answ
\ifx\answ\bigans\message{(This will come out unreduced.}
\magnification=1200\unredoffs\baselineskip=16pt plus 2pt minus 1pt
\hsbody=\hsize \hstitle=\hsize 
\else\message{(This will be reduced.} \let\l@r=L
\magnification=1000\baselineskip=16pt plus 2pt minus 1pt \vsize=7truein
\redoffs \hstitle=8truein\hsbody=4.75truein\fullhsize=10truein\hsize=\hsbody
\output={\ifnum\pageno=0 
  \shipout\vbox{\speclscape{\hsize\fullhsize\makeheadline}
    \hbox to \fullhsize{\hfill\pagebody\hfill}}\advancepageno
  \else
  \almostshipout{\leftline{\vbox{\pagebody\makefootline}}}\advancepageno
  \fi}
\def\almostshipout#1{\if L\l@r \count1=1 \message{[\the\count0.\the\count1]}
      \global\setbox\leftpage=#1 \global\let\l@r=R
 \else \count1=2
  \shipout\vbox{\speclscape{\hsize\fullhsize\makeheadline}
      \hbox to\fullhsize{\box\leftpage\hfil#1}}  \global\let\l@r=L\fi}
\fi
%
\newcount\yearltd\yearltd=\year\advance\yearltd by -1900

\def\Title#1#2{\nopagenumbers\abstractfont\hsize=\hstitle\rightline{#1}%
\vskip 1in\centerline{\titlefont #2}\abstractfont\vskip .5in\pageno=0}
\def\Date#1{\vfill\leftline{#1}\tenpoint\supereject\global\hsize=\hsbody%
\footline={\hss\tenrm\folio\hss}}
%

\def\draftmode{\message{ DRAFTMODE }\def\draftdate{{\rm preliminary draft:
\number\month/\number\day/\number\yearltd\ \ \hourmin}}%
\headline={\hfil\draftdate}\writelabels\baselineskip=20pt plus 2pt minus 2pt
 {\count255=\time\divide\count255 by 60 \xdef\hourmin{\number\count255}
  \multiply\count255 by-60\advance\count255 by\time
  \xdef\hourmin{\hourmin:\ifnum\count255<10 0\fi\the\count255}}}
\def\nolabels{\def\wrlabeL##1{}\def\eqlabeL##1{}\def\reflabeL##1{}}
\def\writelabels{\def\wrlabeL##1{\leavevmode\vadjust{\rlap{\smash%
{\line{{\escapechar=` \hfill\rlap{\sevenrm\hskip.03in\string##1}}}}}}}%
\def\eqlabeL##1{{\escapechar-1\rlap{\sevenrm\hskip.05in\string##1}}}%
\def\reflabeL##1{\noexpand\llap{\noexpand\sevenrm\string\string\string##1}}}
\nolabels
%
\global\newcount\secno \global\secno=0
\global\newcount\meqno \global\meqno=1
\def\newsec#1{\global\advance\secno by1\message{(\the\secno. #1)}
\global\subsecno=0\eqnres@t\noindent{\bf\the\secno. #1}
\writetoca{{\secsym} {#1}}\par\nobreak\medskip\nobreak}
\def\eqnres@t{\xdef\secsym{\the\secno.}\global\meqno=1\bigbreak\bigskip}
\def\sequentialequations{\def\eqnres@t{\bigbreak}}\xdef\secsym{}
\global\newcount\subsecno \global\subsecno=0
\def\subsec#1{\global\advance\subsecno by1\message{(\secsym\the\subsecno. #1)}
\ifnum\lastpenalty>9000\else\bigbreak\fi
\noindent{\it\secsym\the\subsecno. #1}\writetoca{\string\quad
{\secsym\the\subsecno.} {#1}}\par\nobreak\medskip\nobreak}
\def\appendix#1#2{\global\meqno=1\global\subsecno=0\xdef\secsym{\hbox{#1.}}
\bigbreak\bigskip\noindent{\bf Appendix #1. #2}\message{(#1. #2)}
\writetoca{Appendix {#1.} {#2}}\par\nobreak\medskip\nobreak}
%
%
\def\eqnn#1{\xdef #1{(\secsym\the\meqno)}\writedef{#1\leftbracket#1}%
\global\advance\meqno by1\wrlabeL#1}
\def\eqna#1{\xdef #1##1{\hbox{$(\secsym\the\meqno##1)$}}
\writedef{#1\numbersign1\leftbracket#1{\numbersign1}}%
\global\advance\meqno by1\wrlabeL{#1$\{\}$}}
\def\eqn#1#2{\xdef #1{(\secsym\the\meqno)}\writedef{#1\leftbracket#1}%
\global\advance\meqno by1$$#2\eqno#1\eqlabeL#1$$}
%
\newskip\footskip\footskip14pt plus 1pt minus 1pt 
\def\footnotefont{\ninepoint}\def\f@t#1{\footnotefont #1\@foot}
\def\f@@t{\baselineskip\footskip\bgroup\footnotefont\aftergroup\@foot\let\next}
\setbox\strutbox=\hbox{\vrule height9.5pt depth4.5pt width0pt}
\global\newcount\ftno \global\ftno=0
\def\foot{\global\advance\ftno by1\footnote{$^{\the\ftno}$}}
%
\newwrite\ftfile
\def\footend{\def\foot{\global\advance\ftno by1\chardef\wfile=\ftfile
$^{\the\ftno}$\ifnum\ftno=1\immediate\openout\ftfile=foots.tmp\fi%
\immediate\write\ftfile{\noexpand\smallskip%
\noexpand\item{f\the\ftno:\ }\pctsign}\findarg}%
\def\footatend{\vfill\eject\immediate\closeout\ftfile{\parindent=20pt
\centerline{\bf Footnotes}\nobreak\bigskip\input foots.tmp }}}
\def\footatend{}
%
%
\global\newcount\refno \global\refno=1
\newwrite\rfile
\def\ref{[\the\refno]\nref}
\def\nref#1{\xdef#1{[\the\refno]}\writedef{#1\leftbracket#1}%
\ifnum\refno=1\immediate\openout\rfile=refs.tmp\fi
\global\advance\refno by1\chardef\wfile=\rfile\immediate
\write\rfile{\noexpand\item{#1\ }\reflabeL{#1\hskip.31in}\pctsign}\findarg}
\def\findarg#1#{\begingroup\obeylines\newlinechar=`\^^M\pass@rg}
{\obeylines\gdef\pass@rg#1{\writ@line\relax #1^^M\hbox{}^^M}%
\gdef\writ@line#1^^M{\expandafter\toks0\expandafter{\striprel@x #1}%
\edef\next{\the\toks0}\ifx\next\em@rk\let\next=\endgroup\else\ifx\next\empty%
\else\immediate\write\wfile{\the\toks0}\fi\let\next=\writ@line\fi\next\relax}}
\def\striprel@x#1{} \def\em@rk{\hbox{}}
\def\lref{\begingroup\obeylines\lr@f}
\def\lr@f#1#2{\gdef#1{\ref#1{#2}}\endgroup\unskip}

\def\addref#1{\immediate\write\rfile{\noexpand\item{}#1}} 
\def\startrefs#1{\immediate\openout\rfile=refs.tmp\refno=#1}
\def\xref{\expandafter\xr@f}\def\xr@f[#1]{#1}
\def\refs#1{\count255=1[\r@fs #1{\hbox{}}]}
\def\r@fs#1{\ifx\und@fined#1\message{reflabel \string#1 is undefined.}%
\nref#1{need to supply reference \string#1.}\fi%
\vphantom{\hphantom{#1}}\edef\next{#1}\ifx\next\em@rk\def\next{}%
\else\ifx\next#1\ifodd\count255\relax\xref#1\count255=0\fi%
\else#1\count255=1\fi\let\next=\r@fs\fi\next}
%

%
\newwrite\ffile\global\newcount\figno \global\figno=1
\def\fig{fig.~\the\figno\nfig}
\def\nfig#1{\xdef#1{fig.~\the\figno}%
\writedef{#1\leftbracket fig.\noexpand~\the\figno}%
\ifnum\figno=1\immediate\openout\ffile=figs.tmp\fi\chardef\wfile=\ffile%
\immediate\write\ffile{\noexpand\medskip\noexpand\item{Fig.\ \the\figno. }
\reflabeL{#1\hskip.55in}\pctsign}\global\advance\figno by1\findarg}
\def\xfig{\expandafter\xf@g}\def\xf@g fig.\penalty\@M\ {}
\def\figs#1{figs.~\f@gs #1{\hbox{}}}
\def\f@gs#1{\edef\next{#1}\ifx\next\em@rk\def\next{}\else
\ifx\next#1\xfig #1\else#1\fi\let\next=\f@gs\fi\next}
\newwrite\lfile
{\escapechar-1\xdef\pctsign{\string\%}\xdef\leftbracket{\string\{}
\xdef\rightbracket{\string\}}\xdef\numbersign{\string\#}}

\def\writestop{\def\writestoppt{\immediate\write\lfile{\string\pageno%
\the\pageno\string\startrefs\leftbracket\the\refno\rightbracket%
\string\def\string\secsym\leftbracket\secsym\rightbracket%
\string\secno\the\secno\string\meqno\the\meqno}\immediate\closeout\lfile}}
\def\writestoppt{}\def\writedef#1{}
\def\seclab#1{\xdef #1{\the\secno}\writedef{#1\leftbracket#1}\wrlabeL{#1=#1}}
\def\subseclab#1{\xdef #1{\secsym\the\subsecno}%
\writedef{#1\leftbracket#1}\wrlabeL{#1=#1}}
\newwrite\tfile \def\writetoca#1{}
\def\leaderfill{\leaders\hbox to 1em{\hss.\hss}\hfill}
\def\writetoc{\immediate\openout\tfile=toc.tmp
   \def\writetoca##1{{\edef\next{\write\tfile{\noindent ##1
   \string\leaderfill {\noexpand\number\pageno} \par}}\next}}}

%
\catcode`\@=12 
%
\edef\tfontsize{\ifx\answ\bigans scaled\magstep3\else scaled\magstep4\fi}
\font\titlerm=cmr10 \tfontsize \font\titlerms=cmr7 \tfontsize
\font\titlermss=cmr5 \tfontsize \font\titlei=cmmi10 \tfontsize
\font\titleis=cmmi7 \tfontsize \font\titleiss=cmmi5 \tfontsize
\font\titlesy=cmsy10 \tfontsize \font\titlesys=cmsy7 \tfontsize
\font\titlesyss=cmsy5 \tfontsize \font\titleit=cmti10 \tfontsize
\skewchar\titlei='177 \skewchar\titleis='177 \skewchar\titleiss='177
\skewchar\titlesy='60 \skewchar\titlesys='60 \skewchar\titlesyss='60
\def\titlefont{\def\rm{\fam0\titlerm}
\textfont0=\titlerm \scriptfont0=\titlerms \scriptscriptfont0=\titlermss
\textfont1=\titlei \scriptfont1=\titleis \scriptscriptfont1=\titleiss
\textfont2=\titlesy \scriptfont2=\titlesys \scriptscriptfont2=\titlesyss
\textfont\itfam=\titleit \def\it{\fam\itfam\titleit}\rm}
 \ifx\answ\bigans\else scaled\magstep1\fi
\ifx\answ\bigans\def\abstractfont{\tenpoint}\else
\font\abssl=cmsl10 scaled \magstep1
\font\absrm=cmr10 scaled\magstep1 \font\absrms=cmr7 scaled\magstep1
\font\absrmss=cmr5 scaled\magstep1 \font\absi=cmmi10 scaled\magstep1
\font\absis=cmmi7 scaled\magstep1 \font\absiss=cmmi5 scaled\magstep1
\font\abssy=cmsy10 scaled\magstep1 \font\abssys=cmsy7 scaled\magstep1
\font\abssyss=cmsy5 scaled\magstep1 \font\absbf=cmbx10 scaled\magstep1
\skewchar\absi='177 \skewchar\absis='177 \skewchar\absiss='177
\skewchar\abssy='60 \skewchar\abssys='60 \skewchar\abssyss='60
\def\abstractfont{\def\rm{\fam0\absrm}
\textfont0=\absrm \scriptfont0=\absrms \scriptscriptfont0=\absrmss
\textfont1=\absi \scriptfont1=\absis \scriptscriptfont1=\absiss
\textfont2=\abssy \scriptfont2=\abssys \scriptscriptfont2=\abssyss
\textfont\itfam=\bigit \def\it{\fam\itfam\bigit}\def\footnotefont{\tenpoint}%
\textfont\slfam=\abssl \def\sl{\fam\slfam\abssl}%
\textfont\bffam=\absbf \def\bf{\fam\bffam\absbf}\rm}\fi
\def\tenpoint{\def\rm{\fam0\tenrm}
\textfont0=\tenrm \scriptfont0=\sevenrm \scriptscriptfont0=\fiverm
\textfont1=\teni  \scriptfont1=\seveni  \scriptscriptfont1=\fivei
\textfont2=\tensy \scriptfont2=\sevensy \scriptscriptfont2=\fivesy
\textfont\itfam=\tenit \def\it{\fam\itfam\tenit}\def\footnotefont{\ninepoint}%
\textfont\bffam=\tenbf \def\bf{\fam\bffam\tenbf}\def\sl{\fam\slfam\tensl}\rm}
\font\ninerm=cmr9 \font\sixrm=cmr6 \font\ninei=cmmi9 \font\sixi=cmmi6
\font\ninesy=cmsy9 \font\sixsy=cmsy6 \font\ninebf=cmbx9
\font\nineit=cmti9 \font\ninesl=cmsl9 \skewchar\ninei='177
\skewchar\sixi='177 \skewchar\ninesy='60 \skewchar\sixsy='60
\def\ninepoint{\def\rm{\fam0\ninerm}
\textfont0=\ninerm \scriptfont0=\sixrm \scriptscriptfont0=\fiverm
\textfont1=\ninei \scriptfont1=\sixi \scriptscriptfont1=\fivei
\textfont2=\ninesy \scriptfont2=\sixsy \scriptscriptfont2=\fivesy
\textfont\itfam=\ninei \def\it{\fam\itfam\nineit}\def\sl{\fam\slfam\ninesl}%
\textfont\bffam=\ninebf \def\bf{\fam\bffam\ninebf}\rm}
%
%

\hyphenation{anom-aly anom-alies coun-ter-term coun-ter-terms}
\def\inv{^{\raise.15ex\hbox{${\scriptscriptstyle -}$}\kern-.05em 1}}

\def\Dsl{\,\raise.15ex\hbox{/}\mkern-13.5mu D} 
\def\dsl{\raise.15ex\hbox{/}\kern-.57em\partial}

\def\tr{{\rm tr}} 
\font\bigit=cmti10 scaled \magstep1
\def\lspace{\ifx\answ\bigans{}\else\qquad\fi}
\def\lbspace{\ifx\answ\bigans{}\else\hskip-.2in\fi} 
\def\boxeqn#1{\vcenter{\vbox{\hrule\hbox{\vrule\kern3pt\vbox{\kern3pt
	\hbox{${\displaystyle #1}$}\kern3pt}\kern3pt\vrule}\hrule}}}
\def\mbox#1#2{\vcenter{\hrule \hbox{\vrule height#2in
		\kern#1in \vrule} \hrule}}  
%

\def\e#1{{\rm e}^{^{\textstyle#1}}}

\def\darr#1{\raise1.5ex\hbox{$\leftrightarrow$}\mkern-16.5mu #1}

\def\roughly#1{\raise.3ex\hbox{$#1$\kern-.75em\lower1ex\hbox{$\sim$}}}

\def\o{\omega}
\def\s{\sigma}
\def\l{\lambda}
\def\e{{\rm e}}
\def\sqr#1#2{{\vcenter{\vbox{\hrule height.#2pt
     \hbox{\vrule width.#2pt height#1pt \kern#1pt
           \vrule width.#2pt}
       \hrule height.#2pt}}}}

\Title{\vbox{\baselineskip12pt\hbox{USITP-92-09}
\hbox{hep-th/9210081}}}
{{\vbox{\centerline{ A note on  2D chiral gravity and chiral bosons}}}}


\centerline{Fiorenzo Bastianelli\footnote{$^\dagger$}{e-mail:
fiorenzo@vana.physto.se}}
\bigskip\centerline
{\it  Institute for Theoretical Physics}
\centerline {\it University of Stockholm}
\centerline {\it Vanadisv\"agen 9 }
\centerline {\it S-113 46 Stockholm, Sweden}
\vskip 1.3in

\noindent
Quantization of two dimensional chiral matter coupled to gravity
induces an effective action for the zweibein field which is both Weyl and
Lorentz anomalous.
Recently,
the quantization of this induced action  has
been analyzed in the light-cone gauge as well as in the conformal
gauge. An apparent mismatch between the results obtained in the two gauges
is analyzed and resolved by properly treating the Lorentz field as a chiral
boson.

 \vskip .8cm

\Date{09/92}

As is well-known, non-chiral conformal matter in 2d is characterized
by the central charge $c$.
It couples to gravity by inducing  an effective action of the form
$c R {1\over \nabla^2 } R$, which gives a non-trivial dynamics to the
conformal mode of the metric.
Quantization of this induced action
have been performed by fixing the  diffeomorphism invariance
either choosing the light-cone gauge, as done by Polyakov
\ref\P{A. Polyakov, Mod. Phys. Lett. A2 (1987) 893.}
and Knizhnik-Polyakov-Zamolodchikov
\ref\KPZ{ V. Knizhnik,
A. Polyakov and A.B. Zamolodchikov, Mod. Phys. Lett. A3 (1988) 819.},
or choosing the conformal gauge, as shown by David
\ref\Dav{ F. David, Mod. Phys. Lett. A3 (1988) 1651.}
and Distler-Kawai
\ref\DK{ J. Distler and H. Kawai, Nucl. Phys. B231 (1989) 509.}.
In both cases exact results for critical exponents have been obtained,
which agree with each other (when they can
be compared, i.e. at genus zero).
Recently, similar analysis have been attempted for the  case of  gravity
induced by chiral matter
\ref\Yaron{Y. Oz,  J. Pavelczyk and S. Yankielowicz,
Nucl. Phys. B363 (1991) 555.}
\ref\Periwal{R. Myers and V. Periwal, preprint IASSNS-92-19,
Mc-Gill/92-26 and hep-th/9207117,
to appear in Nuclear Physics B.},
the latter being now characterized by the two central charges,
$c_l$ and $c_r$,
appearing in the corresponding left and right Virasoro algebras.
It is well-known that chiral matter
coupled to gravity suffers from gravitational
anomalies
\ref\AG{L. Alvarez-Gaum\'e and E. Witten, Nucl. Phys. B234 (1984)
269.},
which can be conveniently shifted to the Lorentz sector
\ref\Zumino{W. Bardeen and B. Zumino, Nucl. Phys.  B244 (1984) 421.}
by the use of a local counterterm.
With such a choice,
the induced action for the zweibein field
$e_\mu{}^{\pm}$
is  expected to be
diffeomorphism invariant but
Lorentz and Weyl anomalous. It was first computed by
Leutweyler
\ref\Le{H. Leutweyler, Phys. Lett.  B153 (1985) 65.}
for the case of a chiral fermion,
and can be written as follows (see later for notation)
\eqn\effac{
\eqalign{ &\e^{- W[e_\mu{}^{\pm}]} = \int ({\cal D} X) \
\e^{-S[X,e_\mu{}^{\pm}]} \cr
&W[e_\mu{}^{\pm}] = {1\over 24 \pi} \int d^2 x {\sqrt g} \
\biggl ( c_l R_l {1\over \nabla^2} R_l +
c_r R_r {1\over \nabla^2} R_r
+ 2a \o_+ \o_-
\biggr ),\cr}}
where $X$ denotes a generic chiral conformal system.
The parameter $a$  multiplies a local counterterm
which is not fixed by the requirement of diffeomorphism invariance.
For $c_l=c_r =a\equiv c$, this action takes the form
$c R {1\over \nabla^2 } R$, up to boundary terms,
and  Lorentz invariance is recovered.
In ref. \Yaron, such an action was quantized in the light-cone
 gauge for
the zweibein
\eqn\lightcone{ \eqalign{ &e^+ = \e^{-i \lambda} dx^+ \cr
&e^- =
\e^{i\lambda} ( dx^- + h_{++} dx^+) \cr
&\Rightarrow ds^2 = e^+ e^- = dx^+ ( dx^- + h_{++} dx^+),\cr}}
and  an hidden $SL(2,R) \otimes U(1)$ current
algebra was discovered.
 An analysis of the constraints related to such a gauge choice  then
 led the author of ref. \Yaron\
 to a determination of the coefficient
$a$ in term of $c_l$ and $c_r$.
Such a result, however, was not  found in the subsequent
analysis of   Myers and Periwal \Periwal, who chose
to work in the
conformal gauge,
defined by $e_\mu{}^{\pm} = \e^{\s \mp i\lambda} \hat e_\mu{}^{\pm}$.
In this letter, we will address such a discrepancy
by looking at the dynamics of the Lorentz field
before gauge fixing the diffeomorphism invariance.
It will be pointed out that the induced action
describing the dynamics of the Lorentz field coincides with the
action of a  chiral boson, as recently introduced in
\ref\me{F. Bastianelli, Phys. Lett. B227 (1992) 464.}.
This allows us to identify  the stress tensor
of the Lorentz field which enters  the  constraints,
required eventually by the  various gauge fixings.
As a consequence, we do not find any constraint on the value of the coupling
$a$, confirming a conjecture in \Periwal\
according to which the stress tensor of the Lorentz mode was not
correctly identified in ref. \Yaron.

To get started, we
 give our conventions for the two dimensional
local geometry. We introduce Lorentz covariant
 derivatives\footnote
{$^\dagger$}{We work with an euclidean signature and
the flat metric
$\eta_{ab }$
in the complex  coordinates $x^\pm = x^1 \pm i x^2$
is  given by
$\eta_{+- }=\eta_{-+}={1\over2}$ and $ \eta_{++} = \eta_{--}=0$.
$J$ is the Lorentz generator acting on  vectors $V_\pm$
as  $[J,V_\pm] = \pm V_\pm$.}
\eqn\1{ \nabla_a = E_a{}^\mu \partial_\mu + \o_a J,\ \ \ \ \ \ \ \ \ a=\pm}
constrained by
\eqn\2{ \lbrack \nabla_-, \nabla_+ \rbrack = R J.}
The constraint \2 is solved for the spin connection as
($e \equiv det\  e_\mu{}^a$, $e_\mu{}^a$ inverse of $E_a{}^\mu$)
\eqn\3{ \o_\pm = \mp {1\over e} \partial_\mu (e E_\pm{}^\mu ) .}
The definition of the
scalar curvature $R$ then gives
($E_a \equiv E_a{}^\mu \partial_\mu$)
\eqn\R{\eqalign {&R= R_r +R_l\cr
&R_r =\nabla_- \o_+ = (E_- +\o_-) \o_+\cr
&R_l =-\nabla_+ \o_- = -(E_+ -\o_+) \o_-\cr}}
and coincides, up to some normalization, with more standard definitions
which make use of the metric tensor
$g_{\mu\nu} = e_\mu{}^a e_\nu{}^b \eta_{ab} $. Note that with our
definitions the Euler theorem for a surface of genus $g$
reads: ${1\over {\pi}} \int d^2x {\sqrt g} \ R = 2- 2g$.
Another useful object can be defined as $U= R_r - R_l$.
The scaling properties of the various geometrical quantities
under
 Weyl ($\s$) and Lorentz ($\l$) transformations, given by
 $ e_\mu{}^{\pm} \rightarrow \e^{\s \mp i\l}   e_\mu{}^{\pm} $,
are easily derived from the   above formulas, and read
\eqn\tr{ \eqalign{ &\o_{\pm} \rightarrow  \e^{-\s \pm i \l}
\biggl ( \o_\pm \mp \nabla_\pm ( \s \pm i\l) \biggr ), \cr
&e  R_r \rightarrow  e \biggr
( R_r - \nabla^2 (\sigma + i\lambda )\biggl ), \ \ \ \
e  R_l \rightarrow
e \biggr ( R_l - \nabla^2 (\sigma -i\lambda )\biggl ), \cr
&e  R \rightarrow
e \biggr ( R - 2\nabla^2 \sigma \biggl ) , \ \ \ \
e  U \rightarrow
e \biggr ( U- 2i\nabla^2 \lambda \biggl ), \cr}}
where $\nabla^2 \equiv \nabla_- \nabla_+ = \nabla_+ \nabla_-$
 is the laplacian acting on scalars. Note that $R$ is Lorentz invariant
as it should, while $U$ is Weyl invariant.

As already mentioned,
the action induced  by chiral matter has the form in \effac.
Our interest is to quantize such an action, but before fixing
the diffeomorphism invariance by choosing the gauges of refs. \Yaron\ or
 \Periwal, we
want to see how the induced action depends on the Lorentz field.
 We do this by
representing a general zweibein as $e_\mu{}^\pm = \e^{\mp i \l}
\hat e_\mu{}^\pm $, where $\hat e_\mu{}^\pm $ is an arbitrary background.
Dropping the hat on $\hat e_\mu{}^\pm $, we obtain from
the formula in \effac
\eqn\Lorentz{\eqalign{
W[\e^{ \mp i\l }e_\mu{}^{\pm}] &=
W[e_\mu{}^{\pm}] + S_{Lor}[ \l,e_\mu{}^{\pm}] \cr
 S_{Lor}[ \l,e_\mu{}^{\pm}] &=
{1\over 24 \pi} \int d^2 x {\sqrt g} \ \biggl [
\l \nabla^2 \l (2a - c_r - c_l) +2i R_r \l (a - c_r ) -2i R_l \l
( a - c_l ) \biggr ] \cr
&=
{1\over 24 \pi} \int d^2 x {\sqrt g} \ \biggl [
\l \nabla^2 \l (2a - c_r - c_l) +i R \l ( c_l - c_r) +i U \l
(2a - c_r - c_l ) \biggr ] .\cr}}
As anticipated this action is identical to the action for a chiral boson
as described in \me.
Due to the coupling of the  $\l$
field to the background chiral curvature scalars
$R_r$ and $R_l$, the stress tensor acquires improvement terms
which have the effect of shifting asymmetrically the left and
right  central charges.
Note also that
the kinetic term of the Lorentz field is positive definite
for $ c_r +c_l -2a > 0$.
Rescaling the Lorentz field by $ \tilde \l = ({{c_r + c_l -2a }
\over {12}})^{1\over 2} \l$
to get a standard normalization of the kinetic
term\footnote{$^\dagger$}{and standard propagator $\tilde \l (z,\bar z)
\tilde \l(w,\bar w) =
-\log ( | z-w|^2 \mu^2)$,}, we obtain
\eqn\act{ S_{Lor} = {1\over {2 \pi}}
\int  d^2 x {\sqrt g}\ \bigl ( \nabla_+ \tilde \l  \nabla_- \tilde \l
+  2i\beta_r R_r \tilde \l  + 2i\beta_l R_l\tilde \l \bigr )}
with
\eqn\bet{\eqalign{& \beta_l = {c_l -a \over {12}} \biggl (
{12\over {c_l + c_r -2a}} \biggr )^{1\over 2} \cr
& \beta_r =- {c_r -a\over {12}} \biggl (
{12\over {c_l + c_r -2a}} \biggr )^{1\over 2} .\cr}}
The corresponding stress tensor is given by
\eqn\str{\eqalign{
T^{Lor}_{ab} \equiv
{2 \pi \over {\sqrt g}} {\delta S_{Lor} \over \delta e_\mu{}^b} e_{\mu a}
= &
\eta_{ab}E_+ \tilde \l  E_- \tilde \l
-\eta_{+a}E_b \tilde \l  E_- \tilde \l
-\eta_{-a}E_+ \tilde \l  E_b \tilde \l
\cr &+2i\beta_r \bigl (
-\eta_{ab} E_+ E_- \tilde \l
+\eta_{+a}  E_b E_- \tilde \l
+\eta_{-a} \o_+ E_b \tilde \l  \bigr )
\cr &+2i\beta_l \bigl (
-\eta_{ab} E_- E_+ \tilde \l
+\eta_{-a} E_b E_+ \tilde \l
-\eta_{+a} \o_- E_b \tilde \l  \bigr ).
\cr }}
In flat space,
we get the following components of the
energy-momentum tensor
\eqn\T{
\eqalign{
& T^{Lor}_{++}= - {1\over 2} \partial_+ \tilde \l
 \partial_+ \tilde \l
+ i\beta_l \partial_+^2 \tilde \l  ,\ \ \ \
 T^{Lor}_{--}= - {1\over 2} \partial_- \tilde \l
 \partial_-\tilde \l
+ i\beta_r \partial_-^2 \tilde \l ,\cr
& T^{Lor}_{+-}= -i\beta_r \partial_+ \partial_- \tilde \l  =0 , \ \ \ \
 T^{Lor}_{-+} = -i\beta_l \partial_- \partial_+ \tilde \l  =0 ,\cr}}
where  the last  line follows from the
 equation of motion
 $\partial_+ \partial_- \tilde \l
=0$.
Improvement terms have appeared in $T_{\pm\pm}$, turning
the Lorentz field into
a chiral boson. A remarkable  property of the
Lorentz field is that it  contributes
to  the central  charges
in such a way of leveling up
the mismatch between the matter central charges $c_l$ and $c_r$,
independently of the value of $a$,
\eqn\level{ \eqalign{
&c_l^{Lor}= 1- 12 \beta_l^2 =1- {(c_l - a)^2 \over {c_r + c_l -2a}},
 \ \ \ \
c_r^{Lor}= 1-12 \beta_r^2 =1- {(c_r - a)^2 \over {c_r + c_l -2a}}
\cr &\Rightarrow \ \ \ \  c_l + c_l^{Lor} =
c_r + c_r^{Lor} = 1 +  { {c_l c_r - a^2}
\over {c_l + c_r - 2a}}. \cr}}
This  statement can also be checked by integrating
out the Lorentz field in the path integral. Completing squares,
shifting the Lorentz field  and using the
fact that a standard scalar contributes to the induced action by
${1\over {24 \pi}} R {1\over \nabla^2 } R$,
we recover the values in
\level, and, in addition, we can check that $a^{Lor} = 1 -12 \beta_l \beta_r$.
Thus, without the need of introducing additional counterterms,
the induced action due to
the combined effect of the
matter and Lorentz fields reads
\eqn\neweff{
\eqalign{
&W[e_\mu{}^{\pm}] = {{\hat c}\over 24 \pi} \int d^2 x {\sqrt g} \
  R {1\over \nabla^2} R
\cr & \hat c\equiv  c_l + c_l^{Lor} =
c_r + c_r^{Lor} = a + a^{Lor} .\cr}}
Before turning to the various gauge fixing, we note that a vertex operator of
the form $V_k = \exp (ik \tilde \l) $ has  chiral weights
$\Delta_l = {1\over 2} k ( k -2 \beta_l)$ and
$\Delta_r = {1\over 2} k ( k -2 \beta_r)$, corresponding to
 conformal spin $s\equiv \Delta_r - \Delta_l = k
 \bigl (
{ {c_l + c_r -2a}\over {12}} \bigr )^{{1\over 2}}$.

Now we consider the different gauge fixings and
 comment first on the conformal gauge
defined by $e_\mu{}^\pm = \e^\s \hat e_\mu{}^\pm $. We consider the combined
effect of the matter and Lorentz fields, which, as just described,
induce the effective action
for non-chiral gravity with central charge $\hat c$, eq. \neweff.
Everything then proceeds as described in refs. \Dav \DK
\ref\MM{N.E. Mavromatos and  J.L. Miramontes, Mod. Phys. Lett. A4 (1989)
1847.}\ref\DHK{E. D'Hoker and P.S. Kurzepa, Mod. Phys. Lett. A5 (1990)
1441.}. This implies that the Liouville
field $\s$ can be treated as a free field with a central charge
$c^{Liou}= 26 - \hat c$, making the BRST charge
\eqn\BRST{ Q = \oint c(x^+) \biggl ( T_{++}^{matter}+
T_{++}^{Lor}+
T_{++}^{Liou}+
{1\over 2}T_{++}^{gh} \biggr ) (x^+) }
nilpotent, without the need of fixing the coupling $a$
(here $c(x^+)$ is the ghost field of the standard $b,c$
system for the conformal gauge;
a similar formula
applies
to the other chiral sector).

We now switch to the case of the light-cone gauge \lightcone.
The constraint arising from gauge fixing the metric component
$g_{--}=0$ is
\eqn\constr{ T_{++}^{total} \equiv
 T_{++}^{matter}+
T_{++}^{Lor}+
T_{++}^{(h_{++})}
+T_{++}^{gh} =0}
where $T_{++}^{Lor}$ is as in eq. \T,
$T_{++}^{(h_{++})}$ is expressed in term of the $SL(2,R)$ currents and the
the ghost system
contributing to $T_{++}^{gh}$
is the one described in \KPZ.
The requirement that the central charge of $T_{++}^{total}$ vanishes
corresponds to the nilpotency
of the BRST charge for such a gauge choice
and fixes the level of the $SL(2,R)$ current algebra by the equation
\eqn\cc{c_l^{total}= c_l + c_l^{Lor} + {{3K }\over{K+2}} - 6K -28 =0.}
The value of $c_l^{Lor}$ is as in \level.
Thus, no constraint is found on $a$. In ref. \Yaron,
the contribution of the Lorentz field was assumed to be
$c_l^{Lor} =1$, leading through some consistency arguments to a fixed value
of the coupling $a$ (requiring consistency with eq. \level\ would fix
$a=c_l$).
We see that this is not the case.
In fact, in the light-cone gauge we find the following
formulae
\eqn\formula{ \eqalign{ &R=U=R_r=\partial_-^2 h_{++} , \ \ \ \ R_l=0,\cr
&E_+ = \partial_+ -h_{++} \partial_- ,
\ \ \ \  E_- =\partial_- ,
\cr
&\o_+ = \partial_- h_{++} , \ \ \ \  \o_- = 0 ,\cr
}}
and the action in \Lorentz\ takes the form
\eqn\lca{S_{Lor}= {1\over {24 \pi}} \int d^2x \  \biggl [
(2a-c_l -c_r) \lambda \partial_-
(\partial_+ -h_{++} \partial_- )
\lambda +2i (a - c_r ) \lambda \partial_-^2 h_{++} \biggr ].}
Note that because $R=U$ in this gauge, the chiral structure of the induced
action for the Lorentz field
is not anymore manifest.

To conclude, we have analyzed the action for the Lorentz mode of
the zweibein induced by chiral matter and pointed out that
the  Lorentz field behaves as a chiral boson. In particular, we have checked
by computing its stress tensor that it contributes asymmetrically
to the central charges and in such a way of matching the left-right
central charge
asymmetry of the inducing matter.
No constraint is found on the value of the coupling constant $a$ at this
stage of analysis. The formula \level\ indicates the $a$-dependence
of the central charge due to the matter-Lorentz sector.
Our analysis has been of a local nature and we have not dealt with
topology and other global issues such as modular invariance,
which are, of course, essential for a complete understanding of the
chiral gravity model
(see ref. \Periwal\ for steps in this direction),
and might eventually fix a preferred value for $a$.
Also, we have analyzed the Lorentz field
before fixing the gauge for diffeomorphism invariance,
being this of no obstacle  for our purposes. Of course,
other gauges are possible,
where the Lorentz field is not present.
For example, given the general Beltrami parametrization of
the zweibein
\eqn\bel{ \eqalign{ &e^+ = \e^{\s -i \lambda}( dx^+ + h_{--} dx^-)\cr
&e^- =
\e^{\s +i\lambda} ( dx^- + h_{++} dx^+) \cr
&\Rightarrow ds^2 = e^+ e^- =
 \e^{2\s}(dx^+ + h_{--}dx^-)( dx^- + h_{++} dx^+),\cr}}
one could  imagine of fixing the gauge by setting $ \s =\l=0$.
However, the advantage  of the usual light-cone gauge
is lost, since the area element is now non-trivial: ${\sqrt g}
= {\sqrt { 1- h_{++} h_{--}}}$.
Still, the quantum theory of $ h_{++}$ and $  h_{--}$
might shed a different light on the properties of  quantum chiral
gravity.
 \footatend\vfill\supereject\immediate\closeout\rfile\writestoppt
\baselineskip=14pt\centerline{{\bf References}}\bigskip{\frenchspacing%
\parindent=20pt\escapechar=` \input refs.tmp\vfill\eject}\nonfrenchspacing
\end